# DFT investigation of 3d transition metal NMR shielding tensors in diamagnetic systems using the gauge-including projector augmented-wave method

Lionel Truflandier,[*] Michaël Paris and Florent Boucher

Institut des Matériaux Jean Rouxel, UMR 6502, Université de Nantes - CNRS,

2, rue de la Houssinière BP 32229, 44322 Nantes Cedex, France

We present a density functional theory based method for calculating NMR shielding tensors for 3d transition metal nuclei using periodic boundary conditions. Calculations employ the gauge-including projector augmented-wave pseudopotentials method. The effects of ultrasoft pseudopotential and induced approximations on the second-order magnetic response are intensively examined. The reliability and the strength of the approach for $^{49}$Ti and $^{51}$V nuclei is shown by comparison with traditional quantum chemical methods, using benchmarks of finite organometallic systems. Application to infinite systems is validated through comparison to experimental data for the $^{51}$V nucleus in various vanadium oxide based compounds. The successful agreement obtained for isotropic chemical shifts contrasts with full estimation of the shielding tensor eigenvalues, revealing the limitation of pure exchange-correlation functionals compared to their exact-exchange corrected analogues.

---

[*] Electronic addresses: Lionel.Truflandier@cnrs-imn.fr or Florent.Boucher@cnrs-imn.fr

# I. INTRODUCTION

Nuclear magnetic resonance (NMR) spectroscopy is a powerful technique to investigate the structures of molecules, solids or biomolecular systems. For extended systems, the interpretation of spectra provides useful information with regard to the chemical local environment, the number of sites, the coordination number, the internuclear distances or the degree of distortion of polyhedra. In some cases, high resolution NMR measurements can even be used to determine crystallographic space groups.[1] However assignment and interpretation of the resonance lines often remains delicate. This problem can be partially overcome by performing first principles calculations of NMR parameters, i.e. shielding tensors and, for nuclear-spin larger than ½, electric field gradient (EFG) tensors. The development of theoretical methods to calculate NMR properties is currently underway in several scientific communities.[2-5] To perform tractable NMR calculations, one has to deal with the size of the systems under investigation and with the high dependence of the methods with respect to the various levels of approximation, which can significantly affect the computational resources needed. Furthermore, the time-scale for NMR spectroscopy is slow compared with the rovibrational effects of a chemical system. Thus, in order to get quantitative agreement between experimental and calculated results, we have to look beyond static calculations and internal motion contributions to NMR parameters have to be evaluated. Excluding dynamic disorder, those effects can usually be neglected in solid state NMR due to the restricted atomic motion compared with liquid measurement.[1] The reader may find discussions about the state of the art in NMR calculations in several reviews. The review by Helgaker et al.,[4] for instance, gives a broad description of the various quantum chemical methods developed in computational chemistry. The primary effects involved in NMR calculations are described in the de Dios and Facelli reviews.[6,7]

Concerning EFG tensors, it is now well established that they can obtained, at a high level of precision, by performing accurate ground state density calculations. The EFG is directly related to the asphericity of the electron density in the vicinity of the nucleus probe. Various approaches can be used to obtain the full tensor components, the choice being specifically dependant on the type of system under study.[3,8-13] For shielding tensors, the problem is much more complicated. Until recently, the common calculation methods have been based on a molecular approach using localized atomic orbitals (LAOs), the cluster approach being used to mimic infinite periodic systems. However, two important problems remain. Firstly, investigations of molecular materials are carried out by isolating a molecule from the bulk. As a consequence, the chemical environment is neglected in the calculations even though intermolecular interactions may contribute to the shielding and quadrupolar parameters.[14,15] Secondly, in the case of a non-molecular material, the most common compounds in solid state chemistry, strong difficulties of calculations and convergence problem usually occur when using a finite size model. [16]

To overcome such difficulties, Pickard and Mauri have developed the so-called "gauge-including projector augmented-wave" (GIPAW) pseudopotential approach in which the periodicity of the system is explicitly taken into account using a plane-wave basis set to expand the wave functions.[5] This approach was proposed within the framework of density functional perturbation theory (DFPT). The advantage of the GIPAW approach over other pseudopotential methods [17,18] is the possibility of keeping the nodal properties of the wave functions in the neighbourhood of the core in the presence of a magnetic field. Considering the rigid contribution of core electrons with respect to NMR parameters,[19] accuracy comparable to all electron calculations can be achieved.[5] Nevertheless the application to extended systems was, to date, limited to elements belonging to the first three rows of the periodic table,[20-23] due to the difficulties involved in efficient pseudopotential development.

Nowadays, NMR spectroscopy applied to transition metals is widely used in the fields of coordination, bio-chemistry and solid state materials. Among the 3d transition metals, numerous NMR measurements on $^{51}$V nuclei have been performed in order to probe the vanadium(+V) sites in homogeneous and heterogeneous catalysis,[24] battery materials or metalloproteins.[25,26] In this paper we will investigate the calculation of $^{49}$Ti and $^{51}$V NMR shielding tensors in organometallic and diamagnetic inorganic systems, using complexes of titanium and vanadium, and vanadium(+V)-based compounds as representative cases. We will explore for the first time the accuracy of the pseudopotential GIPAW approach on 3d transition metal referring to all electron calculations obtained from traditional quantum chemical methods, the purpose being to apply the computational methodology on extended systems. In Sec. II, we will briefly explore the theoretical methods commonly used in computational chemistry, in order, first, to outline the context in which the GIPAW method was developed, and second to underline approximations and difficulties inherent in the use of a pseudopotential plane-wave method and its application to 3d elements. In Sec. III, we will present the sensitivity of the shielding tensor components accuracies with respect to the level of improvement of the pseudopotential generation. Afterwards, transferability will be checked by means of a benchmark of titanium and vanadium complexes and validated by comparison to all-electron calculations. Application to $^{51}$V containing extended systems will be discussed in Sec. IV. A first example of such an application has been published recently on the AlVO$_4$ system.[27] In this last part, we will finally concentrate on the relation between exchange-correlation functional improvement and reliability of the results.

## II. THEORETICAL METHODS

### A. The electronic current density and the gauge problem

The response of matter to a uniform external magnetic field **B** can be represented by an induced electronic current density **j(r)** which is associated with the operator **J(r)**, through the following relation given in atomic units,

$$\mathbf{J}(\mathbf{r}) = -\frac{1}{2}\{\mathbf{p},|\mathbf{r}\rangle\langle\mathbf{r}|\} - \frac{1}{c}\mathbf{A}(\mathbf{r})|\mathbf{r}\rangle\langle\mathbf{r}|. \tag{1}$$

Here $\{\mathbf{p},|\mathbf{r}\rangle\langle\mathbf{r}|\}$ denotes the anticommutator of the momentum **p** and projection $|\mathbf{r}\rangle\langle\mathbf{r}|$ operators: $\{\mathbf{p},|\mathbf{r}\rangle\langle\mathbf{r}|\} = \mathbf{p}|\mathbf{r}\rangle\langle\mathbf{r}| + |\mathbf{r}\rangle\langle\mathbf{r}|\mathbf{p}$. **A(r)** is a vector potential connected to **B** through $\mathbf{B} = \nabla \times \mathbf{A}(\mathbf{r})$ or $\mathbf{A}(\mathbf{r}) = \frac{1}{2}\mathbf{B} \times (\mathbf{r} - \mathbf{r}_0)$, where **r**$_0$ is the gauge origin. The first and the second parts of the right hand side of the Eq. (1) are the paramagnetic and the diamagnetic current operators, respectively. In a closed shell molecule or insulating non-magnetic material and within the field strengths typically used in NMR experiments, the induced electronic current density is calculated through the first-order-induced current $\mathbf{j}^{(1)}(\mathbf{r})$. It yields a nonuniform induced magnetic field $\mathbf{B}_{in}^{(1)}$ which shields each nucleus N from **B**. The nuclear magnetic shielding tensor $\tilde{\sigma}$ or the so-called chemical shift tensor defined as

$$\mathbf{B}_{in}^{(1)}(\mathbf{r}_N) = -\vec{\vec{\sigma}}(\mathbf{r}_N)\mathbf{B} = \frac{1}{c}\int d^3r\, \mathbf{j}^{(1)}(\mathbf{r}) \times \frac{\mathbf{r}_N - \mathbf{r}}{|\mathbf{r}_N - \mathbf{r}|^3}, \qquad (2)$$

is a second-order magnetic response. The first-order induced current density $\mathbf{j}^{(1)}(\mathbf{r})$ is obtained by means of perturbation theory applied to $\langle \mathbf{J}(\mathbf{r}')\rangle$,[28]

$$\mathbf{j}^{(1)}(\mathbf{r}) = -\sum_o \langle \Psi_o^{(0)} | \{\mathbf{p}, |\mathbf{r}\rangle\langle\mathbf{r}|\} | \Psi_o^{(1)} \rangle - \frac{1}{c}\rho_0 \mathbf{A}(\mathbf{r})$$

$$= \mathbf{j}_p^{(1)}(\mathbf{r}) + \mathbf{j}_d^{(1)}(\mathbf{r}). \qquad (3)$$

In this equation, the summation is over the occupied states $o$ and $\rho_0$ is the unperturbed electron density. The ground state wave function $|\Psi_o^{(0)}\rangle$ is the eigenvector of the field-independant Hamiltonian $\mathbf{H}^{(0)}$ associated with the eigenvalue $\varepsilon_o$ and $|\Psi_o^{(1)}\rangle$ is its corresponding first-order correction due to the magnetic field perturbation. $\mathbf{j}_d^{(1)}(\mathbf{r})$, which depends only on the unperturbed charge density $\rho_0$, is called the "diamagnetic" contribution and corresponds to the uniform circulation of the electrons. $\mathbf{j}_p^{(1)}(\mathbf{r})$, which depends on the first-order perturbed wave function, is called the "paramagnetic" contribution to the total current and is assumed to be a correction due to the molecular environment.

The chemical shift tensor being an observable quantity, $\mathbf{j}^{(1)}(\mathbf{r})$ must be independent of the choice of gauge origin $\mathbf{r}_0$. Both $\mathbf{j}_p^{(1)}(\mathbf{r})$ and $\mathbf{j}_d^{(1)}(\mathbf{r})$ are separately gauge dependent, nevertheless only their sum must satisfy the gauge invariance property. The gauge dependence of $\mathbf{j}_d^{(1)}(\mathbf{r})$ is explicit through the presence of $\mathbf{A}$, while the gauge dependence of $\mathbf{j}_p^{(1)}(\mathbf{r})$ is implicitly present in $\Psi_o^{(1)}$. Different approaches can be used to evaluate $\Psi_o^{(1)}$ such as the Sternheimer equation, the Green's function method, the sum over states approach or the Hylleraas variational principle.[29] All these methods use the first order perturbed Hamiltonian $\mathbf{H}^{(1)}$. For a perturbation due to an external magnetic field,

$$\mathbf{H}^{(1)} = \frac{1}{2c}\mathbf{L} \cdot \mathbf{B}, \qquad (4)$$

where $\mathbf{L} = \mathbf{r} \times \mathbf{p}$ is the angular-momentum operator. Thus, the presence of $\mathbf{H}^{(1)}$ in the calculation of $\Psi_o^{(1)}$ is responsible for the implicit gauge dependence of $\mathbf{j}_p^{(1)}(\mathbf{r})$.

Due to incomplete basis set, the gauge origin independence on $\mathbf{j}^{(1)}(\mathbf{r})$ is usually not completely verified, and it could in principle yield numerical divergence of the calculation of $\mathbf{j}^{(1)}(\mathbf{r})$. Actually, the diamagnetic term converges faster than the paramagnetic part with respect to the basis size. In fact, the diamagnetic term converges quite easily, since only an accurate determination of the ground state density is needed. Considering the paramagnetic contribution, careful choice of gauge origin[30] can lead to a decrease in its magnitude over a particular region of space. As a consequence, a smaller error in the calculated value of $\mathbf{j}^{(1)}(\mathbf{r})$ is expected. The problem of different convergence rates is entirely solved when considering the simple case of an isolated closed-shell atom: $\mathbf{j}_p^{(1)}(\mathbf{r})$ vanishes when the intuitive choice of gauge origin is taken at the nucleus.

Several methods have been developed to solve the gauge problem for molecular systems, using localized atomic orbitals (LAOs). In the limit of complete basis sets, without dependence on the magnetic field, the calculated magnetic shielding tensor should be gauge invariant.[31] Nevertheless, only small molecules have been studied in such a way because of the prohibitive computational effort required.[32,33] An alternative and practical method has been developed through the use of LAOs including explicit field dependence. This well known approach called "gauge invariant atomic orbital" (GIAO) was introduced first by London and generalized for molecular systems by Ditchfield over 30 years ago.[34,35] Each one-electron function has its own local gauge origin represented by a multiplicative complex factor. Latter, Keith and Bader have presented new methods based on the calculation of $\mathbf{j}^{(1)}(\mathbf{r})$ by performing a gauge transformation for each point of space.[36] The "continuous set of gauge transformations" method (CSGT) achieves gauge-invariance via a parametric function $\mathbf{d}(\mathbf{r})$ which is defined in real space and shifts continuously the gauge origin. The potential vector is redefined as,

$$\mathbf{A}(\mathbf{r}) = \frac{1}{2}\mathbf{B} \times (\mathbf{r} - \mathbf{r}_0 - \mathbf{d}(\mathbf{r})).  \tag{5}$$

The type of CSGT method is determined by the choice of the $\mathbf{d}(\mathbf{r})$ function.[2,19,36] If $\mathbf{d}(\mathbf{r})$ is a constant, the single gauge origin method is obtained. In their first work, Keith and Bader proposed a partition of the induced current density into contributions of atoms in a molecule.[30] This method called "individual gauges for atoms in molecules" (IGAIM) is based on the displacement of the gauge origin to the position of the nearest nucleus to the point $\mathbf{r}$ at which $\mathbf{j}^{(1)}(\mathbf{r})$ is calculated. In other words, the function $\mathbf{d}(\mathbf{r})$ takes discrete values equal to the atomic center positions present in the molecule. For chemical shift calculations, CSGT and IGAIM methods give similar results.[36]

GIAO, CSGT and IGAIM methods have been developed for molecular NMR calculations using localized basis sets. The difficulty associated with application of localized methods to extended systems was circumvented by the use of a cluster approximation.[37-42] The accuracy of the results is closely related to the basis quality and the cluster size, and limited convergence was reached despite extensive computational effort. To overcome the difficulties associated with solid state systems, an alternative approach was proposed using the fully periodic GIPAW method.[5]

B. The gauge-including projector augmented-wave approach

In order to discuss the approximations introduced in the magnetic field dependant GIPAW approach, we need first to briefly describe the projector augmented-wave (PAW) electronic structure calculation method elaborated by Blöchl.[12] Within the frozen core approximation and the pseudopotential plane-wave formalism, the PAW method was developed by introducing an operator **T** that maps the true valence wave functions $|\Psi\rangle$ onto pseudo-wave functions $|\widetilde{\Psi}\rangle$, $|\Psi\rangle = \mathbf{T}|\widetilde{\Psi}\rangle$. The construction of **T** is carried out through the use of all-electron (AE) and pseudo (PS) atomic wave functions (so-called AE and PS partial waves), respectively $|\phi_i\rangle$ and $|\widetilde{\phi}_i\rangle$. As in other pseudopotential methods, a cutoff radius $r_{N,c}$ (for each nucleus N) is used to define the augmentation region $\mathbf{\Omega}_N$ where the operator **T** must restore the complete nodal structure of the AE wave functions,

$$\mathbf{T} = \mathbf{1} + \sum_{N,n} \Big( |\phi_{N,n}\rangle - |\widetilde{\phi}_{N,n}\rangle \Big) \langle \widetilde{p}_{N,n}|, \qquad (6)$$

Local projector functions $\langle \widetilde{p}_{N,n}|$ are introduced to expand the pseudo-wave function locally onto the pseudo-atomic orbitals. The index *n* refers to the angular-momentum quantum numbers and to an additional number, which is used if there is more than one projector per angular-momentum channel. Constraints[12] are imposed by the PAW method: $|\widetilde{p}_{N,n}\rangle$ and $|\widetilde{\phi}_{N,n}\rangle$ have to be orthogonal inward $\mathbf{\Omega}_N$ and vanish beyond this region $|\widetilde{p}_{N,n}\rangle$, whereas $|\phi_{N,n}\rangle$ are identical to $|\widetilde{\phi}_{N,n}\rangle$. The evaluation of an observable quantity represented by an operator **O** can be expressed in terms of pseudo-wave functions by $\langle \widetilde{\Psi} | \mathbf{T}^+ \mathbf{OT} | \widetilde{\Psi} \rangle = \langle \Psi | \mathbf{O} | \Psi \rangle$, with an accuracy comparable to an AE calculation. However, within the framework of practical PAW calculations, completeness conditions can not be achieved. The results are dependent on the PS wave function plane-wave basis set expansion and on the AE and PS atomic wave function number.

The ability of the PAW method to reconstruct an AE wave function has allowed the use of the pseudopotential plane-wave formalism for calculations of hyperfine and EFG parameters.[3,43] The efficiency of the EFG calculations has been demonstrated for a large series of nuclei.[44-47] Nevertheless, when considering a second-order magnetic response as the shielding tensor, intricacies appears. It was demonstrated that the PAW approach does not preserve the translational invariance of eigenvectors in the presence of a uniform magnetic field.[5]

The solution proposed by Pickard and Mauri, similar to the GIAO method, is to introduce a field dependant phase factor to the GIPAW method. Here, the multiplicative complex factor is carried out by the operator,

$$\mathbf{T_B} = \mathbf{1} + \sum_{N,n} e^{\frac{i}{2c}\mathbf{r}\cdot\mathbf{r_N}\times\mathbf{B}} \left( |\phi_{N,n}\rangle - |\widetilde{\phi}_{N,n}\rangle \right) \langle \widetilde{p}_{N,n} | e^{-\frac{i}{2c}\mathbf{r}\cdot\mathbf{r_N}\times\mathbf{B}} . \quad (7)$$

As a result, the GIPAW pseudo-eigenvector $|\overline{\Psi}\rangle$ associated with the all-electron-eigenvector $|\Psi\rangle$ is defined by $|\Psi\rangle = \mathbf{T_B}|\overline{\Psi}\rangle$. For a local or semilocal operator, introducing $\xi_N = e^{\frac{i}{2c}\mathbf{r}\cdot\mathbf{r_N}\times\mathbf{B}}$, the GIPAW pseudo-operator $\overline{\mathbf{O}} = \mathbf{T_B^+ O T_B}$ is given by

$$\overline{\mathbf{O}} = \mathbf{O} + \sum_{N,n,m} \xi_N |\widetilde{p}_{N,n}\rangle \left( \langle \phi_{N,n} | \xi_N^+ \mathbf{O} \xi_N | \phi_{N,m} \rangle - \langle \widetilde{\phi}_{N,n} | \xi_N^+ \mathbf{O} \xi_N | \widetilde{\phi}_{N,m} \rangle \right) \langle \widetilde{p}_{N,m} | \xi_N^+ . \quad (8)$$

If one applies the transformation given in Eq. (8) on the operator $\mathbf{J}(\mathbf{r})$ described in Eq. (1), the GIPAW current density operator becomes

$$\overline{\mathbf{J}}(\mathbf{r}') = -\frac{1}{2}\{\mathbf{p}, |\mathbf{r}'\rangle\langle\mathbf{r}'|\} - \frac{1}{c}\mathbf{A}(\mathbf{r}')|\mathbf{r}'\rangle\langle\mathbf{r}'| + \sum_N \xi_N(\mathbf{r})\left(\Delta\mathbf{J}_N^p(\mathbf{r}') + \Delta\mathbf{J}_N^d(\mathbf{r}')\right)\xi_N^+(\mathbf{r}). \quad (9)$$

The GIPAW nodal structure reconstruction leads to the introduction of the paramagnetic $\Delta\mathbf{J}_N^p(\mathbf{r}')$ and diamagnetic $\Delta\mathbf{J}_N^d(\mathbf{r}')$ operators defined in the augmentation region $\Omega_N$;

$$\Delta\mathbf{J}_N^p(\mathbf{r}') = \frac{1}{2}\sum_{n,m}|\tilde{p}_{N,n}\rangle\left(\langle\phi_{N,n}|\{\mathbf{p},|\mathbf{r}'\rangle\langle\mathbf{r}'|\}|\phi_{N,m}\rangle - \langle\tilde{\phi}_{N,n}|\{\mathbf{p},|\mathbf{r}'\rangle\langle\mathbf{r}'|\}|\tilde{\phi}_{N,m}\rangle\right)\langle\tilde{p}_{N,m}|, \quad (10)$$

$$\Delta\mathbf{J}_N^d(\mathbf{r}') = \frac{\mathbf{B}\times(\mathbf{r}'-\mathbf{r}_N)}{2c}\sum_{n,m}|\tilde{p}_{N,n}\rangle\left(\langle\phi_{N,n}|\mathbf{r}'\rangle\langle\mathbf{r}'|\phi_{N,m}\rangle - \langle\tilde{\phi}_{N,n}|\mathbf{r}'\rangle\langle\mathbf{r}'|\tilde{\phi}_{N,m}\rangle\right)\langle\tilde{p}_{N,m}|. \quad (11)$$

If one develops $\overline{\mathbf{J}}$ in powers of $\mathbf{B}$ and uses density-functional perturbation theory,[29] the GIPAW first-order current density is obtained and expressed in different contributions,[5]

$$\mathbf{j}^{(1)}(\mathbf{r}') = \mathbf{j}_{\text{bare}}^{(1)}(\mathbf{r}') + \mathbf{j}_{\Delta p}^{(1)}(\mathbf{r}') + \mathbf{j}_{\Delta d}^{(1)}(\mathbf{r}'), \quad (12)$$

As in Eq. (4) the first order perturbed Hamiltonian is required and expressed thanks to an expansion in powers of $\mathbf{B}$ of the GIPAW pseudo-Hamiltonian $\overline{\mathbf{H}} = \mathbf{T}_B^+ \mathbf{H} \mathbf{T}_B$. Obviously, the expression for $\overline{\mathbf{H}}$ depends entirely on the pseudopotential approach used: Either the norm-conserving[48,49] (NCPP) or the ultrasoft[50] (USPP) schemes. In this latter case, the relaxation of the norm-constraint imposes an additional generalised orthonormality constraint which must be solved via an overlap operator $\mathbf{S}$,

$$\langle\tilde{\phi}_{N,n}|\mathbf{S}|\tilde{\phi}_{N,m}\rangle = \delta_{n,m}. \quad (13)$$

Due to this additional degree of freedom, the simplifications (see Eqs. (11) and (12) with the following discussion of Ref. [5]) which are valid for a NCPP are no longer valid within the USPP-GIPAW approach. The work of Yates has permitted development and implementation of the USPP-GIPAW formalism.[51] Due to the introduction of the generalized orthonormality constraint, the first-order perturbed wave function $\left|\overline{\Psi}_n^{(1)}\right\rangle$ given in Eq. (32) of Ref. [5] is redefined as,

$$\left|\overline{\Psi}_n^{(1)}\right\rangle = \mathbf{G}(\varepsilon_n)(\overline{\mathbf{H}}^{(1)} - \varepsilon_n \mathbf{S}^{(1)})\left|\overline{\Psi}_n^{(0)}\right\rangle, \quad (14)$$

with the Green function operator $\mathbf{G}(\varepsilon_n)$ expressed through

$$\mathbf{G}(\varepsilon_n) = \sum_e \frac{\left|\overline{\Psi}_e^{(0)}\right\rangle\left\langle\overline{\Psi}_e^{(0)}\right|}{\varepsilon_n - \varepsilon_e}, \quad (15)$$

with the sum running over empty states $e$. $\overline{\mathbf{H}}^{(1)}$ and $\mathbf{S}^{(1)}$ are respectively the first-order perturbed GIPAW Hamiltonian and the first-order perturbed overlap matrix. The Green function involving virtual subspace is only used here for convenience in order to express the first-order perturbed wave function $\left|\overline{\Psi}_n^{(1)}\right\rangle$ of Eq. (14). Practically, [51] the closure relation based on the summation of the occupied and virtual subspaces, coupled with a conjugate-gradient minimization scheme leads to a simple linear system of equations, involving solely the occupied ground state wave functions.[52,53] This advantageous scheme, which reduces considerably the computational time, succeeds to express the three different contributions of Eq. (12) as

$$\begin{aligned}\mathbf{j}_{\text{bare}}^{(1)}(\mathbf{r}') = &2\sum_o \text{Re}(\langle\overline{\Psi}_o^{(0)}|\{\mathbf{p},|\mathbf{r}'\rangle\langle\mathbf{r}'|\}\mathbf{G}(\varepsilon_o)(\overline{\mathbf{H}}^{(1)} - \varepsilon_o \mathbf{S}^{(1)})|\overline{\Psi}_o^{(0)}\rangle) - \frac{1}{2c}\rho^{\text{ps}}(\mathbf{r}')\mathbf{B}\times\mathbf{r}' \\ &- \sum_{oo'}\langle\overline{\Psi}_o^{(0)}|\{\mathbf{p},|\mathbf{r}'\rangle\langle\mathbf{r}'|\}|\overline{\Psi}_{o'}^{(0)}\rangle\langle\overline{\Psi}_{o'}^{(0)}|\mathbf{S}^{(1)}|\overline{\Psi}_o^{(0)}\rangle\end{aligned}, \quad (16)$$

where $o$ runs over the occupied states. $\rho^{\text{ps}}(\mathbf{r}') = 2\sum_o \langle\overline{\Psi}_o^{(0)}|\mathbf{r}'\rangle\langle\mathbf{r}'|\overline{\Psi}_o^{(0)}\rangle$ is the ground state pseudo-density. The paramagnetic augmentation current is given by

$$\mathbf{j}_{\Delta p}^{(1)}(\mathbf{r}') = \sum_{N,o} \Bigg( 4\text{Re}(\langle \overline{\Psi}_o^{(0)} | \Delta \mathbf{J}_N^p(\mathbf{r}') G(\varepsilon_o)(\overline{\mathbf{H}}^{(1)} - \varepsilon_o \mathbf{S}^{(1)}) | \overline{\Psi}_o^{(0)} \rangle)$$
$$+ 2\langle \overline{\Psi}_o^{(0)} | \frac{1}{2ic}[\mathbf{B} \times \mathbf{r}_N \cdot \mathbf{r}', \Delta \mathbf{J}_N^p(\mathbf{r}')] | \overline{\Psi}_o^{(0)} \rangle \Bigg) - 2\sum_{oo'} \langle \overline{\Psi}_o^{(0)} | \Delta \mathbf{J}_N^p(\mathbf{r}') | \overline{\Psi}_{o'}^{(0)} \rangle \langle \overline{\Psi}_{o'}^{(0)} | \mathbf{S}^{(1)} | \overline{\Psi}_o^{(0)} \rangle \quad , \quad (17)$$

and the diamagnetic augmentation current is

$$\mathbf{j}_{\Delta d}^{(1)}(\mathbf{r}') = 2\sum_{N,o} \langle \overline{\Psi}_o^{(0)} | \Delta \mathbf{J}_N^d(\mathbf{r}') | \overline{\Psi}_o^{(0)} \rangle \quad . \quad (18)$$

The introduction of extra terms in the expression of $\mathbf{j}_{\text{bare}}^{(1)}(\mathbf{r})$, resulting from the additional orthonormality constraint, yields more awkward calculations compared to the norm-conserving GIPAW method. The NCPP-GIPAW equations can be recovered by putting $\mathbf{S} = \mathbf{1}$ (Eqs. (36) and (37) of Ref. [5]). In order to increase tractability and accuracy of calculations, the gauge origin in the GIPAW approach is put at the nucleus center setting $\mathbf{r}_0 = \mathbf{r}_N$.[19] By reformulating Eqs. (36) and (37) of Ref. [5], it has been shown that the first-order induced current expressed in Eq. (12) is invariant upon a rigid translation through the individual invariance of its three contributions. Then, for a sufficient basis set expansion, the same rate of convergence is observed for $\mathbf{j}_{\text{bare}}^{(1)}(\mathbf{r}')$ and $\mathbf{j}_{\Delta p}^{(1)}(\mathbf{r}')$ (the convergence is governed by the first terms of the right hand sides of Eqs. (16) and (17)). Finally, in order to reduce the computational resources required for the chemical shielding tensor calculations, the first-order induced magnetic field is divided into four contributions which can be individually calculated, taking advantage of the linearity of Eq. (2),

$$\mathbf{B}_{\text{in}}^{(1)}(\mathbf{r}_N) = \mathbf{B}_{\text{core}}(\mathbf{r}_N) + \mathbf{B}_{\text{bare}}^{(1)}(\mathbf{r}_N) + \mathbf{B}_{\Delta p}^{(1)}(\mathbf{r}_N) + \mathbf{B}_{\Delta d}^{(1)}(\mathbf{r}_N) \quad . \quad (19)$$

$\mathbf{B}_{\text{core}}(\mathbf{r}_N)$, which depends only on the core electrons, i.e. of the isolated atom, is calculated once using the Lamb formula.[54]

At this stage, several approximations are introduced to compute NMR chemical shift tensors from the GIPAW approach. Firstly, to evaluate the correction to $\mathbf{B}_{in}^{(1)}(\mathbf{r}_N)$ due to the paramagnetic and diamagnetic augmentation terms, only the augmentation region $\mathbf{\Omega}_N$ of the nucleus N is considered, i.e. the sum on N in Eqs. (16) and (17) is no longer carried out. This on-site approximation neglects the effects of the augmentation currents of the neighbouring atoms to the shielding of the studied atom. Secondly, within periodic conditions, $\mathbf{B}_{in}^{(1)}$ is formulated in reciprocal space using the Biot-Savart law. Unfortunately, for a null vector of the reciprocal lattice ($\mathbf{G} = 0$), $\mathbf{B}_{in}^{(1)}(\mathbf{G}=0)$ becomes a macroscopic quantity.[17] The induced field depends on the surface currents and, as a result, on the shape of the sample. Therefore, the macroscopic magnetic susceptibility $\tilde{\chi}$ has to be evaluated and no full GIPAW approach is available at the moment. Thus, this quantity is calculated using only the $\mathbf{j}_{bare}^{(1)}(\mathbf{r}')$ contribution. Finally, the pseudopotential used for GIPAW calculation must be chosen with caution. Earlier studies show good agreement between all electron (IGAIM) and pseudopotential GIPAW (NCPP) calculations.[5] For a noteworthy reduction of plane-wave expansion, USPP-GIPAW calculations are able to reproduce NCPP-GIPAW results.[51] Without neglecting the intrinsic pseudopotential generation parameters, and especially for 3d elements, the choice of the valence states as well as the number of projectors must be precisely examined in order to reach converged NMR shielding parameters. This issue will be investigated in the next section.

### C. Computational details: all-electron and USPP-GIPAW calculations

In this part we review the default computational parameters employed for the study. If different settings are used, then the calculation details will be explicitly given in the text.

In order to validate the shielding tensor GIPAW calculations for the titanium and vanadium atoms, the USPP-GIPAW results have been compared to those obtained through the AE approach. The Gaussian 03 suite of programs[55] was used to compute all-electron magnetic response of molecules within the IGAIM approach,[30] combined with the "Perdew-Wang 91" exchange and correlation functional PW91.[56,57] Molecular geometries were optimized with symmetry constraints, using the B3LYP hybrid functional[58,59] with the 6-311+G(2d,p) basis set.[60-63] The default force tolerance parameter of 0.02 eV/Å was kept. We considered different kinds of LAOs in order to check the basis set dependence on the shielding tensor calculations of vanadium and titanium atoms. The triple-$\zeta$ 6-311++G(3df,3pd) Pople's basis set developed by Watchers and Hay[62-64] for the first-row transition elements, the augmented triple-$\zeta$ atomic natural orbital (ANO) of Roos and coworkers, tabulated from Sc to Cu atoms,[65] as well as Dunning's quintuple-$\zeta$ correlation-consistent basis set (cc-pCV5Z) developed for the Ti atom by Bauschlicher[66] were used. The basis sets for elements in the first three rows were adapted in order to be consistent with those used for 3d transition metals. For extended systems, all the calculations were carried out using the PW91 functional. The geometry optimization and GIPAW investigations were performed using the CASTEP and NMR-CASTEP codes,[5,20,67] respectively. The Brillouin zone was sampled using Monkhorst-Pack technique.[68] Relaxation of ionic positions were performed at an energy cutoff of 600 eV, using a $k$-point spacing always smaller than 0.05 Å$^{-1}$ and keeping experimental unit cells. The residual forces on atom positions were converged within 0.05 eV/Å. Molecules were studied with 1 $k$-point by the use of a super-cell approach, checking that the super-cell is large enough to avoid spurious interaction between periodic images. This condition was in general satisfied in a 12000 Bohr$^3$ (~12×12×12 Å$^3$) simulation cell. Shielding tensor calculations for molecular and extended systems were carried out through the crystal approach.[5] The interaction of nuclei and core states with the valence electrons was taken into account by the use of USPPs.[50,69] The selection of core levels were the common ones: 1s, [He]2s2p and [Kr]4d for the elements of the second row, third row and for calcium, and lanthanum, respectively. Two projectors were introduced for each remaining ns and np valence states and for the specific case of the 1s valence state of hydrogen two projectors were also used. The core radii $r_c$, beyond which the pseudo-wave functions match the all-electron ones, are given in parentheses (a.u.) for the various atoms: H(0.8), C(1.4), N(1.5), O(1.3), F(1.4), Mg(2.8), P(1.8), S(1.7), Cl(1.7), Ca(1.8) and La(2.3). $r_c$ was set to the same value for all angular momentum channels of a given atom. Moreover, nonlinear core corrections were employed,[70] with a cutoff radius equal to $0.7 \times r_c$. Finally, the same USPP settings were used for DFT geometry optimization as well as shielding parameters calculations, apart from the 3d elements, where the USPP settings for the GIPAW calculations are given explicitly in the text.

### D. Conventions

The conventions used to calculate the chemical shift parameters $\{\delta_{iso}, \delta_{aniso}, \eta_\delta\}$, from chemical shift tensor eigenvalues $\{\delta_{xx}, \delta_{yy}, \delta_{zz}\}$, are defined as follows,

isotropic component:
$$\delta_{iso} = \frac{1}{3}\left(\delta_{xx} + \delta_{yy} + \delta_{zz}\right), \tag{20}$$

anisotropy component:
$$\delta_{aniso} = \delta_{iso} - \delta_{zz}, \tag{21}$$

asymmetry component:
$$\eta_\delta = \frac{\delta_{xx} - \delta_{yy}}{\delta_{aniso}}, \tag{22}$$

with
$$|\delta_{zz} - \delta_{iso}| \geq |\delta_{xx} - \delta_{iso}| \geq |\delta_{yy} - \delta_{iso}|, \qquad (23)$$

The shielding parameters {$\sigma_{iso}$, $\sigma_{aniso}$, $\eta_\sigma$} are deduced from the calculated eigenvalues using relations similar to (18), (19) and (20). One obtains $\sigma_{iso} = \frac{1}{3}(\sigma_{xx}+\sigma_{yy}+\sigma_{zz})$ and $\eta_\sigma = \eta_\delta$ while $\sigma_{aniso} = -\delta_{aniso}$ according to the relation

$$\delta_{ij} = -a \cdot [\sigma_{ij} - \sigma_{ref}] \qquad (24)$$

where $\delta_{ij}$ and $\sigma_{ij}$ are the chemical shift and absolute shielding tensor components respectively, $a$ is a slope (equal to unity in experiments) and $\sigma_{ref}$ is the isotropic shielding of a reference compound. Unfortunately, first-principles calculations of $\sigma_{ref}$ involve the consideration of rovibrational and intermolecular effects. In order to circumvent such tricky calculations, $\sigma_{ref}$ was evaluated assuming a linear regression between computed $\sigma_{iso}$ and experimental $\delta_{iso}$ values.

### III. GIPAW: APPLICATION TO 3d TRANSITION METALS

#### A. Validation of the frozen core approximation

Within the framework of the pseudopotential approximation, the GIPAW method is able to converge towards all-electron magnetic response calculations. One contributing factor of this success is the assumption of a rigid contribution to the shielding NMR parameters of core electrons, i.e. the validity of the frozen core approximation.[17,19,71] The main concept is that the core electrons are not involved in the chemical reactivity, i.e. the core wave functions of an atom remain unmodified whatever its chemical environment is. Therefore, the AE atomic potential can be replaced by a pseudopotential which mimics the potential created by the nucleus surrounded by its inner electrons. The orthogonality condition between the valence and the core states being relaxed, the valence wave functions become smoother and easier to calculate using plane-wave basis sets. For second and third row elements, the core-valence states separation is quite obvious and usual selections of core states are employed by the community for first principle PP calculations. Difficulties appear for the fourth row elements, especially for the 3d transition metals.[72]

Comparing atomic total energies, using the frozen-core PAW and fully relaxed calculations, previous studies have demonstrated that favorable choice of core-valence separation, in terms of computational cost, leads to less accurate results.[73] In the case of the vanadium atom, inaccurate results were found when keeping the {1s2s2p3s3p} states as core states (in the following discussion core and valence shells will be distinguished by the use of braces and parentheses, respectively) while including the 3p states into the valence improved the precision. Consequently, for the first-order magnetic response calculation applied to 3d transition metals through DFPT calculations, one must carefully check the gap between core and valence states.

Within the frozen core approximation and GIAO approach, Schreckenbach and Ziegler have concluded that,[74] for the third period nuclei, the 2p state must be included explicitly in the valence to get accurate results. They also mentioned that for a 3d transition metal like $^{53}$Cr, the 3s and 3p valence shells are necessary. More recently, using the IGAIM approach and choosing the gauge origin at the nucleus center (see Eq. (5)), investigations for $^{29}$Si and $^{31}$P atoms have demonstrated that the core contribution to the chemical shielding is purely diamagnetic,[19] corresponding to a rigid participation of the {1s2s2p} core shells to the shielding tensor. Those contradictory conclusions led us to study the influence of the core-valence partition involved in the GIPAW chemical shielding tensors calculations for 3d elements. We present in Table I the shielding tensor calculated for $^{51}$V in the well-known VOCl$_3$ molecule, using different vanadium pseudopotentials, going from a large {1s2s2p3s3p} to a small {1s2s} core. As previously suggested in the literature,[5] two projectors per channel were used for each angular-momentum, except in the case of the (3s3p4s3d)-GIPAW calculation and for (2p3s3p4s3d)-GIPAW where only one projector is used for the both the 2p and 3s channels. In all cases, the energy cutoff was set large enough to reach convergence for the calculated shielding values with respect to the basis size. A dramatic discrepancy, compared to the AE calculation, is observed in Table I when only the (4s3d) shells are used for the valence. The non-rigid core state contribution of the 3p level is obvious when one compares the (4s3d) and (3p4s3d)-GIPAW calculations. Furthermore, considering the anisotropy parameter, a better agreement between GIPAW and all-electron IGAIM calculations is obtained for an extension of the valence states up to the 3s and even 2p atomic functions. Unfortunately, comparing the $^{51}$V isotropic shielding convergence for the (3s3p4s3d) and (2p3s3p4s3d)-GIPAW calculations (Figure 1), with respect to the cutoff energy, application to solid state systems is not tractable when including the 2p functions in the valence states.

B. Pseudopotential optimization and convergence

In order to demonstrate the computational efficiency of the USPP-GIPAW approach applied to 3d element shielding tensor calculations, we have plotted in Figure 1 the convergence evolution of a NCPP and USPP. For the NCPP case, the core-valence interaction was described by the Troullier-Martin[49] scheme, in the Kleinman-Bylander[75] form. To be consistent with the previous calculation, we used the same core-valence separation and projector allocation as for the (3s3p4s3d)-USPP. The cutoff radii were obviously reduced to a reasonable value of 0.9 au. Moreover, to also demonstrate the interest of using optimized USPP for the vanadium atom,[76,77] we present the convergence results obtained for a non-optimized (3s3p4s3d)-USPP. The method for generating optimized pseudopotentials was introduced by Rappe, Rabe, Kaxiras and Joannopoulos (RRKJ).[76] The RRKJ scheme is based on the statement that, for isolated pseudo-atoms, the total energy convergence is mainly dependant of its kinetic part, which governs the total energy of extended systems. Therefore, to achieve optimal convergence, the authors have proposed a direct method to minimize the high Fourier components of the pseudo-wave functions. Keeping the constraints of normalization and continuity of two derivatives at $r_c$, the pseudization function is optimized in order to minimize the kinetic energy beyond the cutoff wave vector $q_c$. For the non-optimized USPP, using a default value of $q_c = 12.7$ au$^{1/2}$, the $^{51}$V isotropic shielding is converged to within 0.5 ppm at a cutoff energy of 750 eV (Fig. 1). The optimized USPP obtained by setting the $q_c$ parameter to 5.3 au$^{1/2}$, allows reduction of the energy cutoff by about 200 eV. For the same level of accuracy, using NCPP, the cutoff must be dramatically augmented to 3000 eV, which forbids definitively its use for 3d metal shielding calculations involving the (3s3p4s3d) valence states. Finally, whatever the selected GIPAW core-valence separation or pseudopotential scheme are, one should carefully check the convergence using extended basis sets. The same remarks stands for the IGAIM method.

## C. Completeness of the basis set

Within the framework of the PAW method, the completeness of the basis set depends on both the plane-wave energy cutoff and on the AE and PS partial-wave function expansions. With respect to the "additive augmentation principle",[12] Blöchl has shown that the truncation of the partial-wave extension does not affect the completeness of the basis set, assuming the complementary participation of the plane-wave expansion. In order to have a tractable implementation of the PAW formalism for electronic structure calculations, this author has demonstrated that the use of a finite number of partial wave functions yields negligible discrepancy by comparison to AE calculations. To check the transferability of those properties beyond the GIPAW method and to compare shielding parameters with fully converged IGAIM values, we have investigated the convergence of the method with regard to the number of projectors used for each valence state. The validation of the shielding convergence with respect to the plane-wave energy cutoff is quite obvious and has been shown previously in Figure 1. If we rewrite Eq. (19) in terms of the isotropic shielding components, we find

$$\sigma_{iso}(\mathbf{r}_N) = \sigma_{core}(\mathbf{r}_N) + \sigma_{bare}^{G \neq 0}(\mathbf{r}_N) + \sigma_{bare}^{G=0}(\mathbf{r}_N) + \sigma_{\Delta p}(\mathbf{r}_N) + \sigma_{\Delta d}(\mathbf{r}_N). \qquad (25)$$

Clearly, for an isolated molecular system such as the VOCl$_3$ molecule, there are no surface currents (see Sec. II.B.) and the $\sigma_{bare}^{G=0}$ component of Eq. (25) should tend to zero. Thus, the value of this component is a useful tool to check the absence of interactions between periodic images of the molecular system, in the limit of very large super-cells. In our calculations the value was always smaller than 0.5 ppm. Figure 2 shows the projector dependence of the various components of Eq. (25). For all the tested configurations, the plane-wave energy cutoff was set to 700 eV and we used a {1s2s2p}(3s3p4s3d) state configuration for the USPP. As expected, the sensitivity of the paramagnetic correction term is larger than the diamagnetic one, with respect to the number of projectors used. The augmentation of the 4s state with two projectors has no effect on the isotropic shielding component. Indeed, since the paramagnetic augmentation current $\mathbf{j}_{\Delta p}^{(1)}(\mathbf{r'})$ is proportional to the angular momentum (Eqs. (4) and (17)), for a *s* angular momentum, only the bare $\mathbf{j}_{bare}^{(1)}(\mathbf{r})$ (which contains a diamagnetic part) and the diamagnetic augmentation $\mathbf{j}_{\Delta d}^{(1)}(\mathbf{r'})$ terms are dependent of the projector extension. Finally the scattering property of the 4s state is well reproduced with at least one projector. On the other hand, augmentation of the 3p and 3d states leads to strong variations of the isotropic shielding components, especially for the paramagnetic augmentation term. While a deshielding effect is observed for a two-augmented 3p state, a shielding effect is obtained for a two-augmented 3d state. Therefore, this antagonistic effect must be countered by a balanced choice of the number of projectors allocated to the 3p and 3d states. Opposite variations are observed (Figure 2) for the bare term and the diamagnetic augmentation correction expressed in Eq. (25). $\sigma_{bare}^{G \neq 0}$ is slightly affected by the pseudo-partial wave expansion, which yields variations within 2 ppm, against 30 ppm for $\sigma_{\Delta d}$. Furthermore, three projectors are needed to achieve convergence of the paramagnetic augmentation term with respect to the 3p and 3d states. Now, if we compare the fully converged IGAIM and GIPAW results (Table II) for the VOCl$_3$ molecule, fairly good agreement is observed between both series of shielding parameters. In order to improve the reliability of the method for 3d transition metals, the shielding parameters of $^{49}$Ti in the simple TiCl$_3$CH$_3$ molecule are also discussed (Table III). The titanium USPP was built using the same core-valence separation and projector allocation as for the vanadium USPP $3s^P 3p^{2 \times P} 4s^{2 \times P} 3d^{2 \times P}$ (see caption of Table I for details). The cutoff radius was set to 1.8 a.u. for all the angular momentum channels. Concerning isotropic shielding, AE calculations performed with ANO as well as correlation-consistent basis sets agree very well with the GIPAW results, whereas a weak discrepancy of 2 % is observed for the anisotropy parameter.

### D. Pseudopotential transferability: application to organometallic systems

After having demonstrated the accuracy of the USPP-GIPAW method in the calculation of shielding parameters for two molecules, namely $VOCl_3$ for the $^{51}V$ and $TiCl_3CH_3$ for the $^{49}Ti$, it is important to test the transferability of our approach in various electronic and geometric environments. Thus, we have worked with benchmarks of eight V and six Ti based molecular diamagnetic systems. Several all-electron calculations of the $^{51}V$ and $^{49}Ti$ isotropic shielding values have been reported in the literature for organometallic systems.[78-84] Here, we have focused our investigations on the complexes presented in Tables II and III, which have been studied in recent works by Bühl et al..[83] Computation of the NMR shielding parameters within the GIPAW approach was investigated through the use of $3s^P3p^{2\times P}4s^{2\times P}3d^{2\times P}$ and $3s^P3p^{3\times P}4s^{2\times P}3d^{3\times P}$ ultrasoft-pseudopotentials (see caption of Table I) which leads to different convergence levels. As pointed out in Tables II and III and keeping in mind the extended range of the absolute shielding components observed for 3d transition metals, excellent agreement is found between the GIPAW and IGAIM approaches, whatever the level of chosen accuracy. For vanadium isotropic values (Table II), the most important relative discrepancies are observed for the $[V(CO)_6]^-$ and $VF_5$ complexes (6 % and 1 %, respectively, for the first level of accuracy), which may be attributed to the singular electronic environment of the vanadium nucleus. This statement is also true for $TiCl_4$ (Table III), which exhibits a discrepancy of 2 % for the second level of convergence, whereas the isotropic value of $[Ti(CO)_6]^{2-}$ compared to AE calculation remains inferior to 1 %.

A global analysis of our results is given in Table IV which also gather previously published calculations on $^{31}$P, $^{29}$Si and $^{13}$C. [5] Regarding the mean absolute deviations between GIPAW and AE, the differences for the anisotropy parameters are larger than for the isotropic shieldings. In the case of the $3s^P3p^{2\times P}4s^{2\times P}3d^{2\times P}$ USPP, GIPAW and AE isotropic shielding values differ by only 6 ppm which is acceptable for the $^{51}$V atom compared to 1.5 ppm for $^{13}$C and 8.8 ppm for $^{31}$P. The average deviation decreases from 17 to 13 ppm for the anisotropy parameters when we used the $3s^P3p^{3\times P}4s^{3\times P}3d^{3\times P}$ USPP, but unfortunately the value related to shielding parameters increases to 10 ppm. Eventually, if we now assess the percentage of deviation of the $^{51}$V isotropic shielding parameters with respect to the calculated value, a comforting mean value of 0.3 % is found (0.6 % for the second level of convergence), against 0.3 % for $^{29}$Si and 3.2 % for the $^{13}$C. The same conclusions can be drawn for the $^{49}$Ti results, and we remark that the average deviation of the anisotropy parameter is divided by a factor 4 compared to the vanadium value.

In an NMR experiment, we are not directly interested in absolute shielding values but rather in chemical shift parameters with regard to a reference. If we now choose VOCl$_3$ as the reference system, then, using Eq. (24) with $a = 1$, we can calculate GIPAW and IGAIM $^{51}$V chemical shifts. From the values reported in Table III, we found a mean relative discrepancy of 1.6 % and 1.3 % between GIPAW and IGAIM calculations for both the levels of convergence and only 0.8 % between the two GIPAW calculations. This last value drops to 0.2 % when excluding the singular [V(CO)$_6$]$^-$ and VF$_5$ systems. As a result, the $3s^P3p^{2\times P}4s^{2\times P}3d^{2\times P}$ USPP is sufficient to achieve accurate $^{51}$V isotropic chemical shift calculations with a reduced computational effort compared to a $3s^P 3p^{3\times P} 4s^{2\times P} 3d^{3\times P}$ USPP calculation. Furthermore, the calculation time using GIPAW method is of the order of IGAIM with the 6-311++G(3df,3pd) basis set, while it is considerably smaller when more extended basis sets such as cc-pCVXZ or ANO are used. Fast and stable convergence of GIPAW calculations could be a promising alternative compared to time consuming LAO methods in the case of 3d elements. This leads us to consider the plane-wave DFT method as an accurate and efficient approach for the calculation of NMR chemical shift in finite organometallic systems

### E. Relativistic effects

A complete investigation of the relativistic effects on vanadium and titanium shielding tensor calculations is beyond the scope of this paper, but some comments have to be given in order to keep in mind the level of approximation used in the GIPAW method. It will also give some hints to clarify the origin of the differences found between the all-electron IGAIM and the USPP-GIPAW methods. Calculation of the NMR shielding tensor can be separated into two steps: The self consistent field (SCF) procedure which at least leads to the unperturbed Kohn-Sham (KS) eigenvalues and orbitals, and the linear response of these orbitals due to the presence of the magnetic field. Thus, two kinds of relativistic effects are distinguished when calculating the shielding parameters:[85] The indirect term which is associated to the energy and shape modifications of the unperturbed KS orbitals induced by a relativistic SCF procedure,[13] and the direct relativistic effects associated to the use of a relativistic field-dependant Hamiltonian which yields additional terms in the shielding tensor expressions.[86-88] Moreover, these terms can be separated in scalar and spin-orbit coupling parts, depending on the level of approximation used.[87,89] Obviously, for a non-consistent use of methods, i.e. if two different levels of relativistic approximations are used for the SCF and shielding calculations, the analysis and comparison of results should be undertaken with caution.

In our investigations, all-electron calculations are performed with no relativistic approximation, whereas in GIPAW method, introduction of indirect relativistic effects is performed through the pseudopotential approximation. Indeed, the atomic pseudopotentials and wave functions are generated by resolving the scalar relativistic Koelling-Hammond equation.[90] Bouten and co-workers have studied NMR shielding predictions of 3d metal oxide ($MO_4^{n-}$ with M = Cr, Mn, Fe) coupling zero-order regular approximation (ZORA) and GIAO methods.[85,87] They have shown that indirect relativistic effects are from three to four times larger than the direct ones with, on isotropic shieldings, an average magnitude of -63 ppm and 17 ppm for the indirect and direct effects, respectively. However, the indirect contribution does not seem to be rigid with respect to the 3d metal and the considered electronic environment. Therefore, this incomplete insertion of indirect effects could explain the small discrepancies toward the USPP-GIPAW and IGAIM results observed in Tables I and II.

Previous studies combining the ZORA and GIPAW methods[88,91-94] have shown the influence of scalar relativity on $^{77}$Se molecular systems. By taking into account both the direct and indirect effects, an average increase of 69 ppm of the selenium isotropic shielding is observed. However, when calculating a relative chemical shift and comparing to experiments, either using a reference system, or better, by applying a linear regression (Eq. 24), no difference is then found between these two calculations. Similar conclusions can be drawn for the $^{125}$Te, where the relativistic effect is even larger and increases the chemical shielding by about 255 ppm.

As a consequence further work on the influence of GIPAW indirect relativistic effects is necessary, in particular to define the magnitude of the indirect contributions on the shielding parameters, but we are confident that reasonably good results can be obtained for the chemical shift when using the current implementation of the USPP GIPAW method. Investigations on third row elements, especially for $^{49}$Ti and $^{51}$V are in progress.

## IV. APPLICATION TO EXTENDED SYSTEMS

### A. Results and discussion

Having validated, on various molecular systems, the NMR shielding calculations for $^{49}$Ti and $^{51}$V using the USPP-GIPAW method, we will explore now for the first time the accuracy of the pseudopotential approach in calculating the shielding parameters of 3d transition metals in extended systems. We will only focus here on the $^{51}$V nucleus, using a ($3s^P 3p^{2\times P} 4s^{2\times P} 3d^{2\times P}$) USPP for the vanadium (see Sec. III.D.) and an energy cutoff of 700 eV. NMR shielding tensors were calculated for thirteen inorganic vanadium systems, chosen to span a large range of chemical shift for the $^{51}$V. Consequently, a total of eighteen distinct vanadium sites have been investigated. The list of compounds is collected in Table V. Considering previous experimental studies,[95] five different types of vanadium species have been established: orthovanadate[96-98] with almost regular tetrahedral units, pyrovanadate[99] with slightly distorted tetrahedra, metavanadates[100,101] with distorted tetrahedra, vanadates[95,101-103] with distorted octahedra and crystal embedded complexes containing distorted vanadium polyhedra with different surrounded atoms[104] (O and N for $VO_2$[acpy-inh]; O, N and S for VO(OEt)(ONS). A schematic representation of the different structural types and local vanadium environments is shown in Figure 3.

For the eighteen vanadium sites, the correlation between calculated isotropic shielding coefficients and experimental isotropic chemical shifts is shown in Figure 4 and evaluated by a linear least-squares fit according to Eq. (24). This regression displays the good accuracy of the GIPAW method considering the value of the slope -1.047(41) (the ideal value being -1.0) and the correlation coefficient of -0.988. The root mean square deviation of 28 ppm is an indication of the attainable precision for a predictive calculation of isotropic chemical shift in inorganic vanadium based systems. It is also important to note that the fitted $\sigma_{ref}$ value (-1939(59) ppm) is in perfect agreement with the isotropic shielding parameter obtained for $VOCl_3$ using an all-electron calculation (Table I). From this linear regression, the theoretical chemical shift parameters have been calculated for the eighteen vanadium sites and compared to the experimental values (Table V). The larger discrepancies between experimental and theoretical isotropic components, observed for $NH_4VO_3$, $\beta$-$VOPO4$ and VO(OEt)(ONS), can be explained by the strong distortion of the first coordination sphere for the vanadium atom. Moreover, for the special case of VO(OEt)(ONS), the metal atom is located in a quite unusual distorted square pyramid environment formed by one sulfur, one nitrogen and three oxygen atoms.

When many inequivalent sites are present in the same structure, the primary interest is not to predict the isotropic chemical shifts, but instead to assign NMR resonances to the different environments of the probe nucleus. As emphasized in Figure 4, when we focus on a short range of chemical shift (between -1450 to -1350 ppm, for instance), the agreement between calculated and experimental values can be improved by a small adjustment of the $\sigma_{ref}$ value. This has been done in Table V for all the compounds having more than one vanadium site. The results are given between parentheses and allow straightforward assignments of the $^{51}$V resonances in the AlVO$_4$,[23] $\alpha$- and $\beta$-Mg$_2$V$_2$O$_7$, and Ca$_2$V$_2$O$_7$ compounds. With a discrepancy of the order of a few ppm, we are able to discriminate inequivalent vanadium sites exhibiting close isotropic chemical shifts. Unfortunately, the previous conclusions are not transferable to anisotropy and asymmetry parameters. Despite the quite reasonable agreement between experimental and theoretical anisotropy parameters obtained for ortho- and pyrovanadates, huge differences are observed for the other families of vanadium-based compounds. Moreover, the asymmetry parameters are generally poorly reproduced (large experimental deviation could be observed in TABLE V). These disagreements suggest the existence of an indirect relation between the degree of distortion of polyhedra and the theoretical $\delta_{aniso}$ reliability. Finally, especially for high anisotropy values, a significant trend of underestimation of the calculated parameter is revealed. In order to check the overall quality of the correlation between experimental and calculated shielding parameters, and to understand the lack of reliability observed for the calculated $\delta_{aniso}$ and $\eta_\delta$, the eigenvalues of the chemical shift tensor have to be considered.[27] Experimental eigenvalues have been obtained from chemical shift parameters using Eqs. (20) to (22), whereas theoretical values have been deduced from absolute shielding eigenvalues, using Eq. (24) and the linear regression previously fitted. We have shown that the classification of chemical shift eigenvalues according to the relation (22) can lead to inversions of calculated components with regard to the experimental values.[27] In order to have a consistent comparison, incorrect assignments have been corrected when needed. The correlation is plotted in Figure 5. When all the eighteen vanadium sites are considered, poor agreement is observed between the experimental and theoretical eigenvalues, which contrast with the very good correlation observed for $\delta_{iso}$ values (Figure 4). This contrasting behavior may be related to an error compensation induced by the averaging process bound to the isotropic values. Looking more carefully at the different classes of compounds, orthovanadates, which reveals low $\delta_{aniso}$ and high $\eta_\delta$ values, are characterized by a low dispersion of the eigenvalues and exhibit certainly the best agreement (see discussion Ref. [27]). This is in contrast to the pyrovanadates, where experimental and theoretical $\delta_{aniso}$ display quite good agreement, yet strong disparities are graphically observed due to the poor reproduction of the $\eta_\delta$ values (Table V). For the other families, the correlation in Figure 4 is even worse, in relation with the increasing polyhedron distortion.

B. Improvement of DFT calculations

The previous results demonstrate the difficulties to quantitatively reproduce the chemical shift components using DFT. Observed discrepancies between experiment and theory are masked by the average isotropic chemical shift and magnified by the anisotropy and asymmetry parameters. Moreover, difficulties in calculating, in specific cases, isotropic chemical shift for $^{17}O$ have been recently reported and discussed for solid state NMR. [105] The authors have invoked the "band gap error" and the inaccuracy of the local density approximation (LDA) and generalized gradient approximation (GGA) exchange-correlation (XC) functionals to properly describe excited state spectra. Other investigations on molecular systems have shown that calculated shielding parameters are highly dependant on the type of XC functionals.[106-110] Linear response of crystalline or molecular orbitals to the magnetic field perturbation are strongly dependant on the occupied-virtual energy gap ($E_{gap}$) and the shape of the virtual orbitals, through the first-order corrected wave function (Eq. 14). Recent studies have shown that hybrid density functionals, which include a portion of Hartree-Fock (HF) exchange, partially overcome the "band gap error" problem in solid state systems.[111-115] In the case of quantum chemical NMR calculations, it was established that implementation of exact exchange in functionals leads to a huge improvement of calculated transition metal isotropic shieldings in organometallic systems.[116] To our knowledge, apart from an isolated computational investigation of the effect of the XC functionals on anisotropy for nuclei in organic molecules,[108] theoretical investigations have mainly been carried out considering the average isotropic component obtained from the three eigenvalues of the second rank shielding tensor.

To discuss the influence of the HF exchange on anisotropy and asymmetry parameters, we now focus our attention on the VOCl$_3$ inorganic system (bulk-optimized geometry have been kept, see Sec. II.C.). Shielding calculations were performed through the use of IGAIM method coupled with the 6-311++G(3df,3pd) basis set. Investigation of the influence of the exact exchange on shielding parameters has been performed using different exchange-correlation functionals. For GGAs, we have used the "Perdew-Wang 91" exchange and correlation functional PW91,[56,57] and the BLYP functional, which combined the "Becke's 1988" exchange and the "Lee-Yang-Parr" correlation functionals.[58,117] Hybrid XC functionals are defined by the following exchange-correlation approximation,

$$E_{XC}^{hybrid} = \alpha E_X^{HF} + (1-\alpha)E_X^{LDA} + \beta \Delta E_X^{GGA} + E_C^{GGA}. \qquad (26)$$

Where $E_X^{HF}$ is the "exact" HF exchange, $E_X^{LDA}$ is the LDA exchange, $\Delta E_X^{GGA}$ and $E_C^{GGA}$ are respectively the exchange correction and correlation parts of GGA functional. We use the three-parameter B3 exchange functional defined by Becke,[59] leading to a value of $\alpha =$ 0.2. The correlation GGA functionals $E_C^{GGA}$ are taken as the Perdew-Wang 91,[56,57] and Lee-Yang-Parr.[58] Results are collected in Table VI. Firstly, in order to probe the packing effect on the $^{51}$V shielding parameters, we have used the cluster approximation using ten additional VOCl$_3$ entities which mimic the bulk environment on a central molecule (Table VI). This procedure works pretty well in the present case if we compare the GIPAW calculations and the IGAIM-cluster results, and validates both approaches. By isolating a unique VOCl$_3$ molecule and comparing to the cluster results, we conclude that the influence of the Van der Waals interactions on calculated shielding parameters are negligible. Thus, calculations carried out with an isolated molecule should be reliable enough to be extrapolated to the fully periodic GIPAW calculations. Inspection of Table VI reveals that the two GGAs as well as the two hybrid functionals give similar results. The differences between both sets of pure and hybrid functionals are around 35 ppm for $\sigma_{iso}$ and 20 ppm for $\delta_{aniso}$. Considering a GGA and the corresponding hybrid functional, we observe a fairly good improvement of $\delta_{aniso}$ with regard to experiment (Table V) when exact exchange is introduced. Afterwards, we have studied the dependence of the calculated shielding eigenvalues on the amount of exact exchange involved in the hybrid functional. This has been done using the half-and-half functional proposed by Becke,[118] and defined with the following relation,

$$E_{XC}^{HandH} = \alpha E_X^{HF} + (1-\alpha)E_X^{LDA} + E_C^{LYP}. \qquad (27)$$

Evolution of the occupied-virtual gap and shielding eigenvalues with regard to the mixing coefficient $\alpha$ are displayed in Figure 6. Increase of the exact exchange leads to a linear widening of the occupied-virtual energy splitting. $E_{gap}$ discrepancy between pure DFT exchange ($\alpha = 0$, called HandH$^0$), and quasi-full HF exchange ($\alpha = 0.9$, called HandH$^{0.9}$) is about 0.30 a.u.. Calculation using Hartree-Fock level of theory (results not shown) gives a value of 0.50 a.u. compared to 0.19 and 0.41 for HandH$^0$ and HandH$^{0.9}$. These results agree with the well-known LDA-GGA underestimation and Hartree-Fock overestimation of occupied-virtual energy gap. Considering the shielding components results, we observed that $\sigma_{iso}$ and $\delta_{aniso}$ are strongly dependent on the exact exchange, and the anisotropy parameter is the more affected. Following the above observations, we could suspect that the anisotropy improvement is closely bound to the correction of the occupied-virtual energy gap induced by the use of hybrid XC functionals.

Nevertheless, according to an extensive study of the influence of pure exchange on $^{57}$Fe isotropic shielding through GIAO-DFT calculations,[107] Schreckenbach has demonstrated that three factors are responsible for the improvement induced by the use of hybrid functionals: enhancement of the occupied-virtual gap, increase of the diffuse character of virtual molecular orbitals and the coupling contribution due to the HF exchange (Eq. (21) from Ref. [107]). All these contributions, and especially the last two, have an important effect on the paramagnetic part of the shielding tensor. As a result further work is in progress to understand quantitatively the influences of the exact exchange on the shielding tensor eigenvalues. At least we can deduce that the discrepancies found for the $^{51}$V anisotropy and asymmetry NMR parameters are probably linked to a fundamental DFT deficiency rather than GIPAW built-in approximations.

## V. CONCLUSION

We have shown that extension of the GIPAW method to 3d nuclei in finite and infinite systems is reliable and reproduces with high accuracy the NMR isotropic shieldings of

$^{51}$V and $^{49}$Ti in diamagnetic molecular-like and extended inorganic systems. The stable and fast convergence of the pseudopotential method is able to overcome difficulties due to the incomplete expansion of the localized basis, reducing considerably the computational cost associated with traditional quantum chemical methods. Moreover the use of scalar relativistic pseudopotentials leads to the introduction of indirect relativistic corrections without increasing calculation time, which are the dominant contribution in 3d transition metals compared to fully relativistic calculations. Furthermore, direct assignment of $^{51}$V solid state NMR resonances is allowed. We have demonstrated that principal components of the shielding tensors should be considered in order to avoid erroneous conclusions on the quality of the theoretical model, when looking for correlation between calculated and experimental results. Despite a lack of reliability observed for anisotropy and asymmetry parameters, we are hopeful that future investigations will correct these limitations of DFT. Finally, we believe that this new approach will be a complementary and useful tool for experimental NMR research applied to organometallic and solid state chemistry.

## ACKNOWLEDGEMENTS

The calculations presented in this work have been carried out at the Centre Régional de Calcul Intensif des Pays de la Loire financed by the French Research Ministry, the Région Pays de la Loire, and Nantes University. L.T. gratefully acknowledges C.J. Pickard for useful discussions and J.R. Yates for providing his PhD thesis manuscript. We also wish to thank C. Payen, N. Dupré and C. Ewels for careful reading of the manuscript.

TABLE I. Convergence of the $^{51}$V absolute isotropic and anisotropic shielding parameters as a function of the vanadium valence states involved in USPP-GIPAW calculations for the VOCl$_3$ molecule. The multi-projector USPP is defined by the notation nl$^{k\times P}$ where an integer k is associated to each nl atomic state and displays the number of projectors allocated (one projector is allocated to the 3s channel).

| Valence State | $r_c^c$ | $v_{loc}^c$ | Number of projectors | $\sigma_{iso}$ (ppm) | $\delta_\sigma$ (ppm) |
|---|---|---|---|---|---|
| (4s3d) | 2.4 | p(-0.5) | 4s$^{2\times P}$3d$^{2\times P}$ | -1806 | -353 |
| (3p4s3d) | 2.5 | f(0.0) | 3p$^{2\times P}$4s$^{2\times P}$3d$^{2\times P}$ | -1910 | -434 |
| (3s3p4s3d) | 2.0 | f(0.0) | 3s$^P$3p$^{2\times P}$4s$^{2\times P}$3d$^{2\times P}$ | -1910 | -455 |
| (2p3s3p4s3d)$^a$ | 0.8/2.0 | f(0.0) | 2p$^P$3s$^P$3p$^{2\times P}$4s$^{2\times P}$3d$^{2\times P}$ | -1920 | -461 |
| all-electron$^b$ | - | - | - | -1904 | -483 |

$^a$ A core radius of 0.8 and 2.0 a.u. was used for the 2p and for the remaining states respectively.

$^b$ IGAIM/6-311++G(3df,3pd).

$^c$ Core radius $r_c$ and atomic reference energies (in parentheses) of the local atomic pseudopotential $v_{loc}$ are given in a.u..

TABLE II. $^{51}$V NMR shielding parameters in various molecular systems. The GIPAW calculations were performed using the (1) $3s^P3p^{2\times P}4s^{2\times P}3d^{2\times P}$ and (2) $3s^P3p^{3\times P}4s^{2\times P}3d^{3\times P}$ ultrasoft pseudopotentials and compared to the IGAIM calculations performed with the (1) 6-311++G(3df,3pd) and (2) ANO-3ζ LAO basis sets.

| $^{51}$V | | $\sigma_{iso}$ (ppm) | | $\delta_\sigma$ (ppm) | | $\eta_\delta$ | |
|---|---|---|---|---|---|---|---|
| Molecule | USPP/LAO | GIPAW | IGAIM | GIPAW | IGAIM | GIPAW | IGAIM |
| VOCl$_3$ | (1) | -1910 | -1904 | -455 | -483 | 0.00 | 0.00 |
|  | (2) | -1947 | -1952 | -463 | -464 | 0.00 | 0.00 |
| [V(CO)$_6$]$^-$ | (1) | 97 | 91 | 0 | 0 | n/a | n/a |
|  | (2) | 89 | 76 | 0 | 0 | n/a | n/a |
| VF$_5$ | (1) | -1220 | -1233 | -9 | -6 | 0.01 | 0.00 |
|  | (2) | -1280 | -1258 | 1 | 11 | 0.14 | 0.00 |
| VOF$_3$ | (1) | -1177 | -1177 | 336 | 317 | 0.00 | 0.00 |
|  | (2) | -1212 | -1214 | 335 | 348 | 0.00 | 0.00 |
| VOClF$_2$ | (1) | -1415 | -1418 | 293 | 290 | 0.37 | 0.27 |
|  | (2) | -1451 | -1458 | 293 | 297 | 0.36 | 0.35 |
| VON$^a$ | (1) | -1546 | -1548 | -42 | -61 | 0.02 | 0.01 |
|  | (2) | -1584 | – | -46 | – | 0.02 | – |
| VOCl$_2$F | (1) | -1663 | -1663 | -345 | -358 | 0.46 | 0.47 |
|  | (2) | -1700 | -1707 | -349 | -347 | 0.46 | 0.45 |
| VO(CH$_3$)$_3$ | (1) | -3034 | -3020 | -1641 | -1652 | 0.00 | 0.00 |
|  | (2) | -3074 | -3057 | -1647 | -1615 | 0.00 | 0.00 |

$^a$ Abbreviation for the VO(OCH$_2$CH$_2$)$_3$N complex. Computation of the NMR shielding parameters was not tractable for this molecule using the ANO-3ζ basis set.

TABLE III. $^{49}$Ti NMR shielding parameters in various molecular systems. The GIPAW calculations were performed using the (1) $3s^P3p^{2\times P}4s^{2\times P}3d^{2\times P}$ and (2,3) $3s^P3p^{3\times P}4s^{2\times P}3d^{3\times P}$ ultrasoft pseudopotentials and compared to the IGAIM calculations performed with the (1) 6-311++G(3df,3pd), (2) ANO-3ζ and (3) cc-pCV5Z LAO basis sets.

| $^{49}$Ti | | $\sigma_{iso}$ (ppm) | | $\delta_\sigma$ (ppm) | | $\eta_\delta$ | |
|---|---|---|---|---|---|---|---|
| Molecule | USPP/LAO | GIPAW | IGAIM | GIPAW | IGAIM | GIPAW | IGAIM |
| TiCl$_3$CH$_3$ | (1) | -1459 | -1471 | 465 | 462 | 0.00 | 0.00 |
|  | (2) | -1491 | -1494 | 471 | 479 | 0.00 | 0.00 |
|  | (3) |  | -1489 |  | 476 |  | 0.00 |
| [Ti(CO)$_6$]$^{2-}$ | (1) | 623 | 621 | 0 | 0 | n/a | n/a |
|  | (2) | 626 | 622 | 0 | 0 | n/a | n/a |
|  | (3) |  | 622 |  | 0 |  | n/a |
| TiCl(CH$_3$)$_3$ | (1) | -2171 | -2183 | -445 | -441 | 0.00 | 0.00 |
|  | (2) | -2206 | -2191 | -451 | -448 | 0.00 | 0.00 |
|  | (3) |  | -2208 |  | -451 |  | 0.00 |
| TiCl$_2$(CH$_3$)$_2$ | (1) | -1845 | -1859 | -483 | -477 | 0.78 | 0.80 |
|  | (2) | -1879 | -1885 | -489 | -491 | 0.78 | 0.80 |
|  | (3) |  | -1876 |  | -487 |  | 0.80 |
| Ti(CH$_3$)$_4$ | (1) | -2434 | -2448 | 0 | 0 | n/a | n/a |
|  | (2) | -2473 | -2468 | 0 | 0 | n/a | n/a |
|  | (3) |  | -2451 |  | 0 |  | n/a |
| TiCl$_4$ | (1) | -778 | -780 | 0 | 0 | n/a | n/a |
|  | (2) | -796 | -780 | 0 | 0 | n/a | n/a |
|  | (3) |  | -781 |  | 0 |  | n/a |

TABLE IV. Comparison between GIPAW and IGAIM methods for various nuclei, using benchmarks of molecules, through the consideration of the deviation $\Delta$ and relative mean absolute deviation $\Delta_r$. The GIPAW $^{51}$V and $^{49}$Ti NMR results were computed using (1) $3s^P3p^{2\times P}4s^{2\times P}3d^{2\times P}$ and (2) $3s^P3p^{3\times P}4s^{2\times P}3d^{3\times P}$ ultrasoft pseudopotentials.

| Nucleus | GIPAW PP-PW | IGAIM LAO basis set | $\Delta^a$ (ppm) $\sigma_{iso}$ | $\delta_\sigma$ | $\Delta_r^b$ (%) $\sigma_{iso}$ |
|---|---|---|---|---|---|
| $^{51}$V | (1) | 6-311++G(3df,3pd) | 5.9 | 16.9 | 0.3 |
|  | (2) | ANO-3$\zeta$ | 10.4 | 13.3 | 0.6 |
| $^{49}$Ti | (1) | 6-311++G(3df,3pd) | 9.4 | 2.9 | 0.6 |
|  | (2) | ANO-3$\zeta$ | 9.5 | 2.7 | 0.7 |
| $^{31}$P$^c$ | Ref. [5] | cc-pCVQZ | 8.8 | - | 2.6 |
| $^{29}$Si$^c$ | Ref. [5] | cc-pCVQZ | 0.8 | - | 0.3 |
| $^{13}$C$^c$ | Ref. [5] | cc-pCVQZ | 1.5 | - | 3.2 |

$^a$ Mean absolute deviation calculated using $\Delta x = \frac{1}{n}\sum_i^n x_i^{IGAIM} - x_i^{GIPAW}$, where $x$ and $n$ are the shielding parameters and the number of molecules respectively. The VO(OCH$_2$CH$_2$)$_3$N molecule was dismissed from the statistic calculation.

$^b$ Relative mean absolute deviation calculated using $\Delta_r x = \frac{1}{n}\sum_i^n \left|\frac{x_i^{IGAIM} - x_i^{GIPAW}}{x_i^{IGAIM}}\right| \times 100$.

$^c$ Calculations were performed using norm-conserving pseudopotential with the LDA exchange-correlation functional. $\Delta$ and $\Delta_r$ calculations related to the $\{^{31}$P, $^{29}$Si, $^{13}$C$\}$ nuclei were accomplished with $n = \{3, 7, 5\}$, from Ref. [5].

TABLE V. Experimental and calculated $^{51}$V shielding parameters ($\delta_{iso}$, $\delta_{aniso}$, $\eta_\delta$) using the USPP-GIPAW method, in various vanadate compounds.

| compound | site | Theoretical (ppm) | | | Experimental (ppm) | | | Ref. |
|---|---|---|---|---|---|---|---|---|
| | | $\delta_{iso}$ | $\delta_{aniso}$ | $\eta_\delta$ | $\delta_{iso}$ | $\delta_{aniso}$ | $\eta_\delta$ | |
| orthovanadate | | | | | | | | |
| AlVO$_4$ | V(1) | -705$^a$(-738$^b$) | -96 | 0.55 | -744 ± 1 | -120 ± 6 | 0.72 ± 0.10 | [97] |
| | V(2) | -633(-670) | -77 | 0.86 | -661 ± 1 | 87 ± 8 | 0.74 ± 0.17 | |
| | V(3) | -742(-773) | -62 | 0.50 | -776 ± 1 | -82 ± 7 | 0.88 ± 0.11 | |
| LaVO$_4$ | | -616$^a$ | -49 | 0.65 | -605 ± 1 | -50 ± 5 | 0.71 ± 0.05 | [98] |
| pyrovanadate | | | | | | | | |
| $\alpha$-Mg$_2$V$_2$O$_7$ | V(1) | -628(-603) | -73 | 0.89 | -604 ± 1 | 103 ± 2 | 0.34 ± 0.16 | [99] |
| | V(2) | -570(-549) | -73 | 0.53 | -549 ± 1 | -57 ± 3 | 0.91 ± 0.10 | |
| $\beta$-Mg$_2$V$_2$O$_7$ | V(1) | -669(-639) | -113 | 0.49 | -639 ± 1 | -113 ± 7 | 0.90 ± 0.10 | [99] |
| | V(2) | -517(-495) | -264 | 0.26 | -494 ± 1 | -262 ± 3 | 0.10 ± 0.10 | |
| Ca$_2$V$_2$O$_7$ | V(1) | -576(-570) | 72 | 0.36 | -575 ± 1 | 71 ± 3 | 0.54 ± 0.35 | [99] |
| | V(2) | -543(-539) | 473 | 0.62 | -534 ± 1 | 530 ± 10 | 0.50 ± 0.03 | |
| metavanadate | | | | | | | | |
| NH$_4$VO$_3$ | | -601 | 156 | 0.37 | -570 ± 1 | 240 ± 5 | 0.70 ± 0.03 | [100] |
| Mg(VO$_3$)$_2$ | | -544 | 263 | 0.21 | -534 ± 1 | 310 ± 3 | 0.30 ± 0.03 | [101] |
| Ca(VO$_3$)$_2$ | | -567 | 414 | 0.39 | -563 ± 1 | 517 ± 5 | 0.18 ± 0.03 | [101] |
| vanadate | | | | | | | | |
| V$_2$O$_5$ | | -622 | 468 | 0.07 | -612 ± 1 | 645 ± 1 | 0.11 ± 0.05 | [102] |
| $\beta$-VOPO4 | | -718 | 484 | 0.01 | -755 | 818 | 0.00 | [103] |
| VOCl$_3$ (103 K) | | 5 | -429 | 0.03 | 7 | -323 | 0.03 | [95] |
| complexe | | | | | | | | |
| VO(OEt)(ONS) | | -310 | 271 | 0.90 | -369 ± 1 | 336 ± 68 | 0.35 ± 0.10 | [104] |
| VO$_2$[acpy-inh] | | -519 | 371 | 0.45 | -504 ± 2 | 485 ± 29 | 0.25 ± 0.25 | [104] |

$^a$ Predictive $^{51}$V chemical shifts have been calculated with respect to the Eq. (24), using $a$ = 1.047 and $\sigma_{ref}$ = -1939.

$^b$ Relative $^{51}$V chemical shifts for AlVO$_4$, $\alpha$- and $\beta$-Mg$_2$V$_2$O$_7$, and Ca$_2$V$_2$O$_7$ are reported relative to the reference values –2004, -1943, -1940 and –1959 respectively ($a$ = 1 in Eq. (24)).

TABLE VI. Influence of the XC functional on the calculated $^{51}$V shielding parameters in VOCl$_3$.

| method | XC functional | $\sigma_{iso}$ (ppm) | $\delta_{aniso}$ (ppm) | $\eta_\delta$ |
|---|---|---|---|---|
| GIPAW | PW91 | -1944 | -429 | 0.03 |
| cluster[a] | PW91 | -1941 | -445 | 0.03 |
| molecule | PW91 | -1924 | -415 | 0.00 |
| molecule | BLYP | -1959 | -434 | 0.00 |
| molecule | B3PW91 | -2185 | -366 | 0.00 |
| molecule | B3LYP | -2226 | -390 | 0.00 |

[a] A cluster of eleven VOCl$_3$ entities have been used, keeping the geometry used for the GIPAW periodic NMR calculations. Shielding parameters are related to the central molecule.

Fig. 1. GIPAW method convergence using different vanadium pseudopotentials (see the Sec. II.A. & B. for the pseudopotential setting details). Calculated $^{51}$V isotropic shielding in VOCl$_3$ molecule is plotted versus the plane-wave energy cutoff $E_c$.

(Color online) Fig. 2. Evolution of the $^{51}$V isotropic shielding components as a function of the number of projectors used in the USPP-GIPAW calculation, for the VOCl$_3$ molecule. The scale of $\sigma_{\Delta p}$ was reduced by a factor of 15 compared to $\sigma_{bare}^{G \neq 0}$ and $\sigma_{\Delta d}$.

(Color online) Fig. 3. Polyhedral projection of the various classes of vanadium-based inorganic systems using representative compounds of Table V: (a) LaVO$_4$ for orthovanadate, (b) $\beta$-Mg$_2$V$_2$O$_7$ for pyrovanadate, (c) NH$_4$VO$_3$ for metavanadate, (d) and (e) represent the vanadate class with $\beta$-VOPO$_4$ and CaVO$_3$ and (f) an organometallic complexe with VO(OEt)(ONS). Structural distortions are shown in terms of distance (given in Å) with their first coordination sphere.

Fig. 4. Plot of the $^{51}$V GIPAW absolute isotropic shielding versus experimental chemical shifts for the 18 vanadium sites referenced in the TABLE V. The solid line represents the linear correlation. All the fitted parameters are given in the upper right panel.

Fig. 5. Experimental versus calculated $^{51}$V chemical shift tensor eigenvalues for the various vanadate compounds of TABLE V. The solid line represents perfect agreement between calculation and experiment.

Fig. 6. Evolutions of the occupied-virtual energy gap and $^{51}$V shielding tensor eigenvalues as a function of the Hartree-Fock mixing coefficient involved in the HandH hybrid exchange-correlation functional for VOCl$_3$.

Figure 1

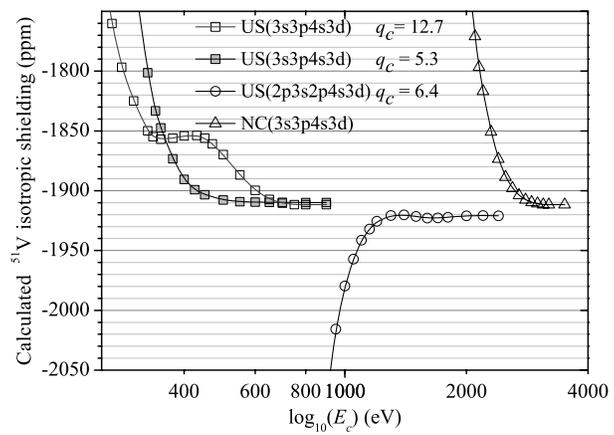

Figure 2

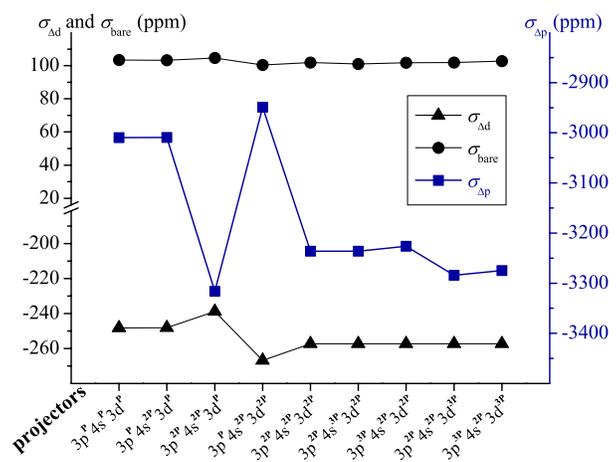

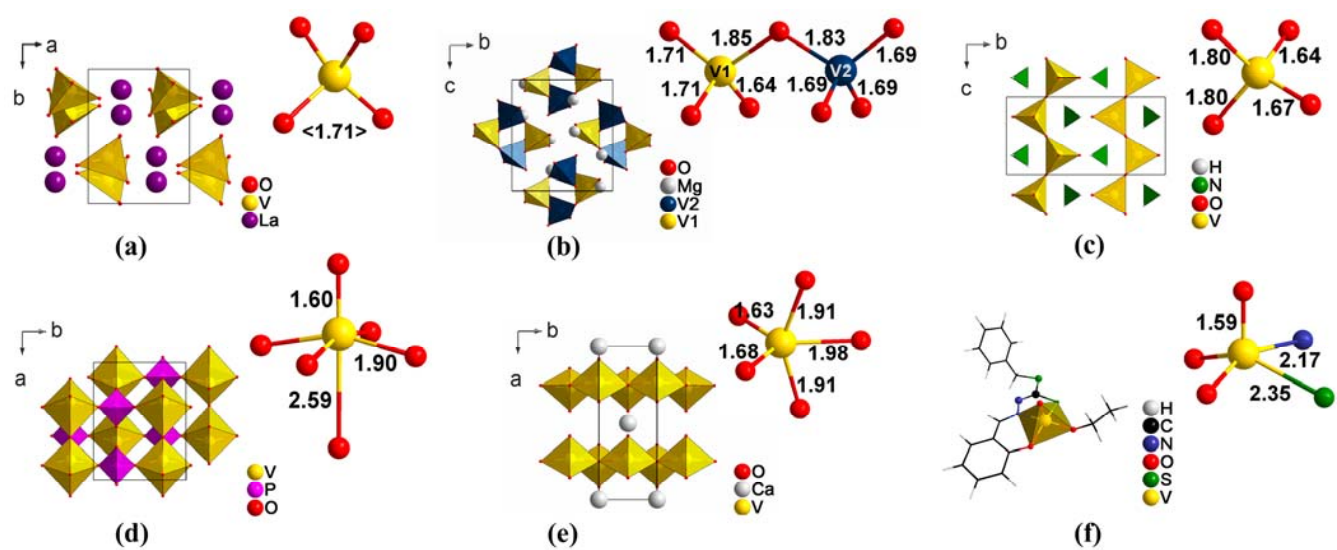

Figure 4

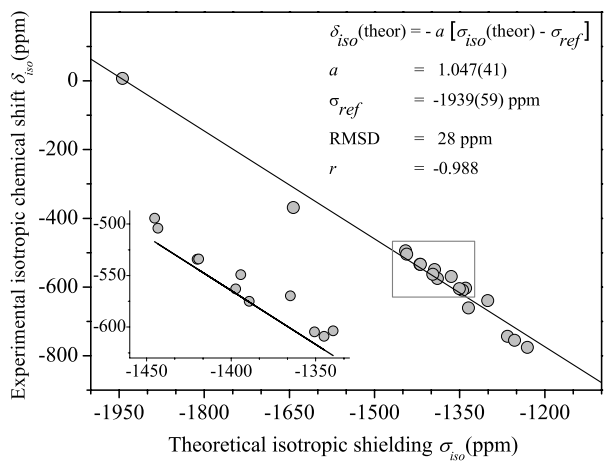

Figure 5

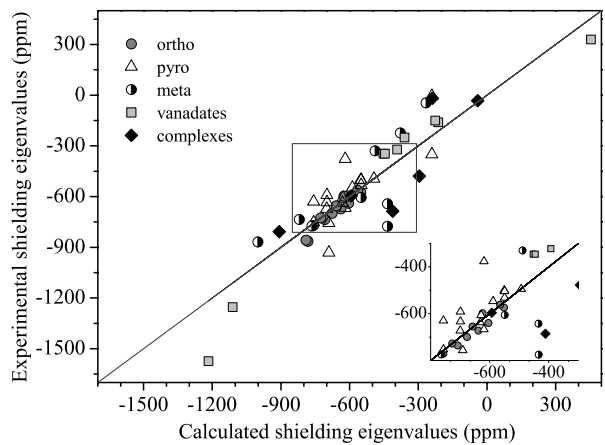

Figure 6

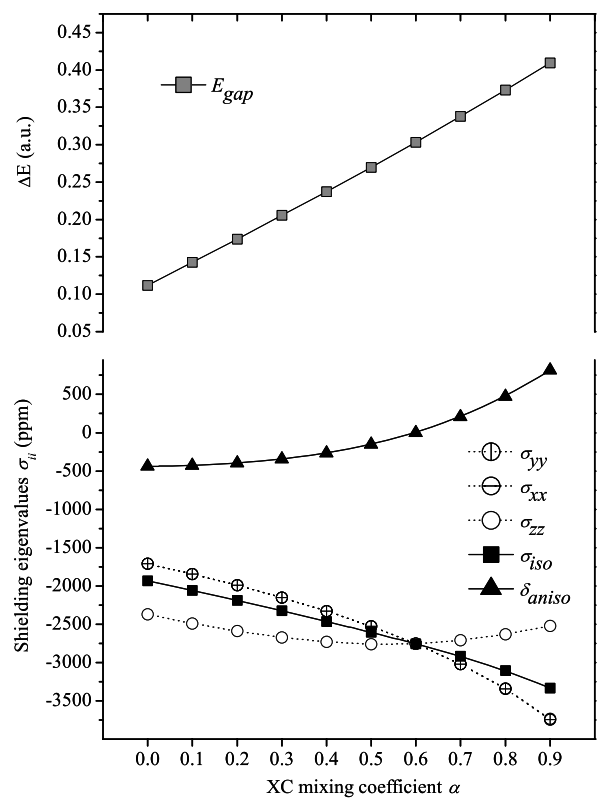